\documentclass{aa}  
\usepackage{graphicx}
\usepackage[english]{babel}
\usepackage{txfonts}
\newcommand{\text}{\mbox}
\newcommand{\vi}{\mathsf{V}_I}
\newcommand{\vq}{\mathsf{V}_Q}
\newcommand{\vu}{\mathsf{V}_U}
\newcommand{\vv}{\mathsf{V}_V}

\renewcommand{\dagger}{\star}

\begin{document}
   \title{An efficient phase-shifting scheme for bolometric additive interferometry}

   \subtitle{}

   \author{R. Charlassier  \and J.-Ch. Hamilton \and \'{E}. Br\'eelle \and  A. Ghribi \and Y. Giraud-H\'{e}raud  \and J. Kaplan \and M. Piat \and D. Pr\^ele}

   \offprints{\tt rcharlas@apc.univ-paris7.fr}

   \institute{APC, Universit\'e Denis Diderot-Paris 7, CNRS/IN2P3, CEA, Observatoire de Paris ; 10 rue A. Domon \& L. Duquet, Paris, France}

   \date{Received ; accepted }

% \abstract{}{}{}{}{} 
% 5 {} token are mandatory
 
  \abstract
   {Most upcoming CMB polarization experiments will use direct imaging to search for the primordial gravitational waves through the B-modes. Bolometric interferometry is an appealing alternative to direct imaging that combines the advantages of interferometry in terms of systematic effects handling and those of bolometric detectors in terms of sensitivity.}
   {We calculate the signal from a bolometric interferometer in order to investigate its sensitivity  to the Stokes parameters paying particular attention to the choice of the phase-shifting scheme applied to the input channels in order to modulate the signal.}
   {The signal is expressed as a linear combination of the Stokes parameter visibilities whose coefficients are functions of the phase-shifts.}
   {We show that the signal to noise ratio on the reconstructed visibilities can be maximized provided the fact that the phase-shifting scheme is chosen in a particular way called "coherent summation of equivalent baselines". As a result, a bolometric interferometer is competitive with an imager having the same number of horns, but only if the coherent summation of equivalent baselines is performed. We confirm our calculations using a Monte-Carlo simulation. We also discuss the impact of the uncertainties on the relative calibration between bolometers and propose a way to avoid this systematic effect.}
   {}


   \maketitle
%
%________________________________________________________________

\section*{Introduction}
Measuring precisely the polarization of the Cosmic Microwave Background (CMB) is one of the major challenges of contemporary observationnal cosmology. It has already led to spectacular results concerning the cosmological model~[\cite{DASI, CBI, WMAPa, WMAPb, QUAD}] describing our Universe. Even more challenging is the detection of the so-called B-modes in the CMB polarization, associated with pure tensor modes originating from primordial gravitational waves enhanced by inflation. Discovering these modes would give direct information on inflation as the amplitude of the B-modes is proportional to the tensor to scalar ratio for the amplitude of the primordial density perturbations which is a direct product of inflationary scenarii~[\cite{refBmodesinflation}]. Furthermore, it seems that most of the inflationary models arising in the context of string theory (brane inflation, ...) predict an undetectably small scalar to tensor ratio~[\cite{KalloshBmodes}]. The discovery of B-modes in the CMB may therefore appear as the only present way to falsify string theories. Cosmic strings and other topological defects are also sources of density perturbations of both scalar and tensor nature. They are however largely dominated by the adiabatic inflationary perturbations in TT, TE and EE power spectra and therefore hard to detect. It is only in the B-mode sector (BB power spectrum) that the tensor topological defects perturbation could be large~[\cite{BevisBmodes}] and have a different shape~[\cite{UrrestillaBmodes}] from those originating from inflation and hence be detectable~[\cite{PogosianBmodes}]. 

Unfortunately, the inflationary tensor to scalar ratio seems to be rather small so that the B-modes are expected at a low level as compared to the E-modes. The quest for the B-modes is a therefore tremendous experimental challenge: one requires exquisitely sensitive detectors with an unprecedented control of the instrumental systematics, observing at a number of different frequencies to be able to remove foreground contamination. Various teams have decided to join the quest, most of them with instrumental designs based on the imager concept (BICEP, EBEX, QUIET, SPIDER, CLOVER). Another possible instrumental concept is an interferometer that has many advantages from the point of view of systematic effects (no optics for instance) and that directly measures the Fourier modes of the sky. Let us recall that the first detections of polarization of the CMB were performed with interferometers~[\cite{DASI,CBI}]. 
Interferometers are however often considered as less sensitive than imagers mainly because of the additional noise induced by the amplifiers required for heterodyne interferometry whereas imagers use background limited bolometers. Another drawback of heterodyne interferometry is that it requires a number of correlators that scales as the square of the number of input channels limiting the number of channels actually achievable~[\cite{cmbtaskforce}].

A new concept of instrument called "Bolometric Interferometer" is currently under developpement (MBI [\cite{MBI}], BRAIN [\cite{Brain, BrainRomain}]). In such an instrument, the interference fringes are "imaged" using bolometers. We believe that such an instrument could combine the advantages of interferometry in terms of systematic effects and data analysis and those of bolometers in terms of sensitivity. The goal of this article is to investigate ways to reconstruct the Fourier modes on the sky (the so-called {\em visibilities}) of the Stokes parameters with a bolometric interferometer. In particular, we focus our attention on the necessary phase-shifting schemes required to modulate the fringe patterns observed with the bolometer array. We show that one can construct phase-sequences that allow to achieve an excellent sensitivity on the visibilitites: scaling as  $\sqrt{N_h}/N_\mathrm{eq}$ (where $N_h$ is the number of horns and $N_\mathrm{eq}$ is the number of couples of horns separated by identical vectors hereafter called {\em equivalent baselines}) whereas it would scale as $\sqrt{N_h}/\sqrt{N_\mathrm{eq}}$ for a non optimal phase-shifting sequence.

This article is organised as follows: in section~\ref{design} we describe the assumptions that we make on the hardware design and on the properties of the various parts of the detector. In section~\ref{stokes} we describe how the signal measured by such an instrument can be expressed in terms of the Stokes parameter visibilities. We show how to invert the problem in an optimal way in section~\ref{reco} and show how the phase-shifting scheme can be chosen so that the reconstruction is indeed optimal in section~\ref{phases}. We have validated the method we propose using a Monte-Carlo simulation described in section~\ref{mc}. We end up by some considerations about systematic effects induced by cross-calibration errors and propose a way to avoid them in section~\ref{syste}.

\section{Bolometric Interferometer design\label{design}}
\begin{figure}[!t]
\centering\resizebox{5cm}{!}{\centering{
\includegraphics[angle=270]{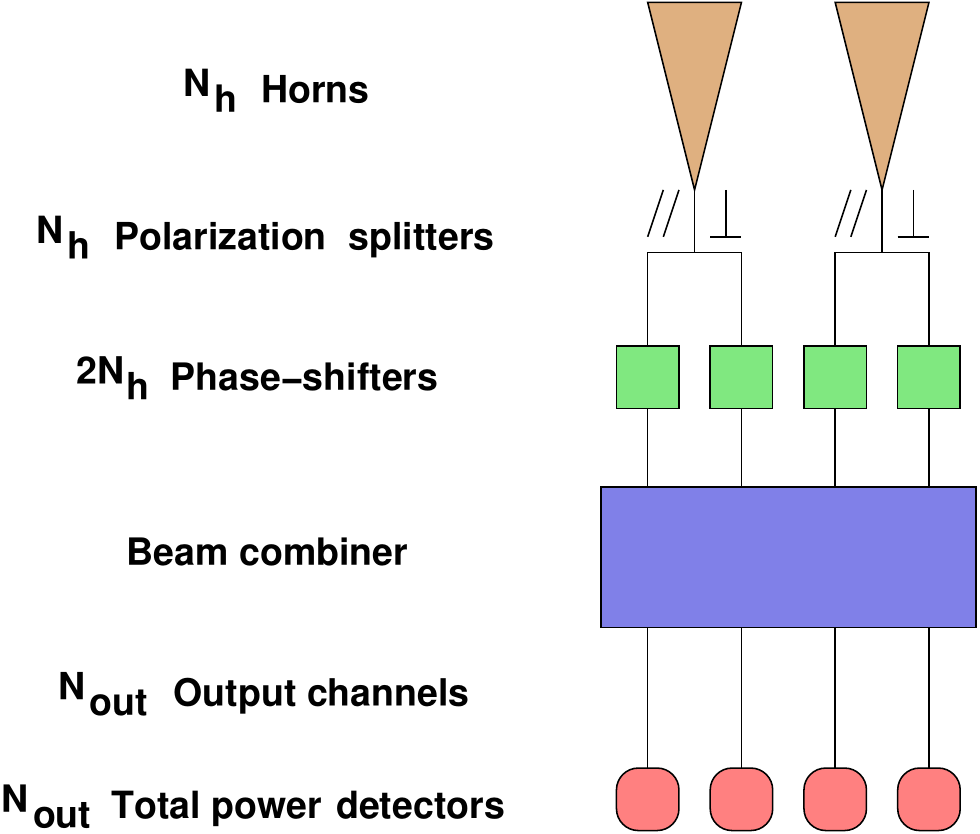}}}
\caption{\small Schematic view of the bolometric interferometer design considered in this article.}
\label{designfig} 
\end{figure}

In this section we will describe the basic design we assume for the bolometric interferometer and how the incoming radiation is transmitted through all of its elements. This will lead us to a model of the signal that is actually detected at the output of the interferometer. A schematic view of the bolometric interferometer is shown in Fig.~\ref{designfig}

 \subsection{Horns}

We assume that we are dealing with an instrument which is observing
the sky through $N_{h}$ input horns placed on an array at positions
$\vec{d}_{i}$. All horns are supposed to be coplanar and  looking
towards the same direction on the sky. They are characterized by their
beam pattern on the sky noted $B_{\mathrm{in}}(\vec{n})$ where $\vec{n}$
is the unit vector on the sphere. Two horns $i$ and $j$ form a baseline
which we label by $0\leq b\leq N_{h}(N_{h}-1)/2-1$. The phase difference
between the signal reaching the two horns from the same direction
$\vec{n}$ of the sky is such that:
\begin{equation}
E_{j}(\vec{n})=E_{i}(\vec{n})\exp(2i\pi\vec{u}_{b}\cdot\vec{n}),\mbox{ where }\vec{u}_{b}=(\vec{d_{j}}-\vec{d}_{i})/\lambda,\label{eq:marchhdiff0}
\end{equation}
where $\lambda$ is the central observing wavelength.

\subsection{Equivalent baselines}

It is clear that if two baselines $b$ and $b'$ are such that $\vec{u}_{b}=\vec{u}_{b'}$,
then the phase shifts associated with the two baselines are equal,
a fact that we shall extensively use in the following. All baselines
$b$ such that $\vec{u}_{b}=\vec{u}_{\beta}$ form a class of equivalent
baselines associated with mode $\vec{u}_{\beta}$ in visibility space.
For all baselines $b$ belonging to the same class $\beta$, the phase
difference between the two horns $i$ and $j$ is the same:
\begin{equation}
E_{j}(\vec{n})=E_{i}(\vec{n})\exp(2i\pi\vec{u}_{\beta}\cdot\vec{n})\label{marchdiff}.
\end{equation}
The number $N_{\neq}$ of different classes of equivalent baselines
depends on the array, and the number of different baselines in an
equivalence class also depends on the particular class. For instance,
if we consider a square array with $N_{h}=N_{\mathrm{side}}^{2}$horns,
there are $N_{\neq}=2\, N_{\mathrm{side}}(N_{\mathrm{side}}-1)$ classes,
and the number of equivalent baselines in the class associated with%
\footnote{in units of the smallest baseline in the array.%
} \begin{eqnarray*}
\vec{u}_{\beta} & = & \left(\begin{array}{c}
l\\
m\end{array}\right)\mbox{ with }1\leq l\leq N_{\mathrm{side}}-1\mbox{ for }m=0\\
 & \mbox{ and } & -|N_{\mbox{side}}-1|\leq l\leq N_{\mathrm{side}}-1\mbox{ for }1\leq m\leq N_{\mathrm{side}}-1,\end{eqnarray*}
is $N_{eq}(\beta)=(N_{\mathrm{side}}-|l|)(N_{\mathrm{side}}-m)$.

\subsection{Polarization splitters}

In order to be sensitive to the polarization of the incoming radiation,
we also assume that at the output of each horn there is a device which
separates the radiation into two orthogonal components noted $\parallel$
and $\perp$. Such a separation can be achieved with an OrthoMode
Transducer (OMT) in wave-guide~{[}\cite{OMTwave}], finline~{[}\cite{OMTfinline}]
or planar~{[}\cite{OMTplanar,OMTplanar2}] technologies. Each horn
therefore has two outputs measuring the electric field integrated
through the beam in the two orthogonal directions. The contribution
coming from direction $\vec{n}$ for polarization $\eta$ ($\parallel$
or $\perp$) is: 
\begin{equation}
\epsilon_{i}^{\eta}(\vec{n})=B_{\mathrm{in}}(\vec{n})E_{i}^{\eta}(\vec{n}).
\end{equation}

\subsection{Phase-shifters}

Important components of the required setup are the {\em phase-shifters}
placed on each of the outputs that allow the phase of the electric
field to be shifted by a given angle that can be chosen and controlled
externally. This is the way the signal is modulated in order to recover
the incoming information. For now we do not make any assumptions on
the possible values of the angles but we will see that they have to
be chosen carefully in order to optimize the signal to noise ratio.
The signal after phase-shifting coming from direction $\vec{n}$ with
polarization $\eta$ is: 
\begin{equation}
\epsilon_{i}^{\prime\eta}(\vec{n})=\exp(i\phi_{i}^{\eta})\epsilon_{i}^{\eta}(\vec{n}).
\end{equation}
 For obvious hardware reasons, all phase-shifters in the setup have
to be identical and deliver the same possible phase-shifts.

\subsection{Beam combiner}

In order to be able to perform interferometry, the beam of each horn
has to be combined with all the others so that all possible baselines
are formed. The realization of a beam combiner is an issue in itself
that will not be assessed in the present article. As an example, this
can be achieved using a Butler combiner~{[}\cite{thesebutler}] or
with a quasi-optical Fizeau combiner such as the one used for
the MBI instrument~{[}\cite{MBI}]. All of these devices are such
that the $2N_{h}$ input channels result after passing through the
beam combiner in $N_{\mathrm{out}}$ output channels that are linear
combinations of the input ones. To be able to conserve the input power
in an ideal lossless device, the number of output channels $N_{\mathrm{out}}$
has to be at least equal to the number of input channels $2N_{h}$. In the
output channel $k$ the electric field is: 
\begin{equation}
z_{k}(\vec{n})=\frac{1}{\sqrt{N_{\mathrm{out}}}}\sum_{i=0}^{N_{h}-1}\sum_{\eta=0}^{1}\gamma_{k,i}^{\eta}\epsilon_{i}^{\eta}(\vec{n})\exp(i\phi_{i}^{\eta}),
\end{equation}
 where the $\gamma_{k,i}^{\eta}$ coefficient model the beam combiner,
$\eta=1$ or $0$ respectively corresponds to $\parallel$ and $\perp$
polarizations. We choose to deal with configurations where the incoming
power is equally distributed among all output channels, meaning that
the coefficients $\gamma_{k,i}^{\eta}$ have a unit modulus: $\left|\gamma_{k,i}^{\eta}(k)\right|=1$
. In order to simplify the notation, we include the $\gamma_{k,i}^{\eta}$
phases in the phase-shifting terms as $\Phi_{k,i}^{\eta}=\phi_{i}^{\eta}+\mathrm{Arg}(\gamma_{k,i}^{\eta})$
so that: \begin{equation}
z_{k}(\vec{n})=\frac{1}{\sqrt{N_{\mathrm{out}}}}\sum_{i=0}^{N_{h}-1}\sum_{\eta=0}^{1}\epsilon_{i}^{\eta}(\vec{n})\exp(i\Phi_{k,i}^{\eta})\label{phaseshift}.
\end{equation}

\subsection{Total power detector}

The signal from each of the outputs of the combiner is not detected
in a coherent way as in a heterodyne interferometer but with a bolometer
through its total power averaged on time scales given by the time
constant of the detector (larger than the frequency of the EM wave).
We assume that the bolometers are background limited, meaning that
their noise variance is proportional to their input power. The power
on a given bolometer is then: 
\begin{eqnarray}
\mathcal{S}_{k} & = & \left<\left|\int z_{k}(\vec{n})\mathrm{d}\vec{n}\right|^{2}\right>_{\mathrm{time}}\\
 & = & \int\left<z_{k}(\vec{n})z_{k}^{\star}(\vec{n}^{\prime})\right>_{\mathrm{time}}\mathrm{d}\vec{n}\mathrm{d}\vec{n}^{\prime}.
 \end{eqnarray}
The signal coming from different directions in the sky are incoherent
so that their time averaged correlation vanishes: 
\begin{eqnarray}
\left<z_{k}(\vec{n})z_{k}^{\star}(\vec{n}^{\prime})\right>_{\mathrm{time}} & = & \left<\left|z_{k}(\vec{n})\right|^{2}\right>_{\mathrm{time}}\delta(\vec{n}-\vec{n}^{\prime})\\
 & \equiv & \left|z_{k}(\vec{n})\right|^{2}\delta(\vec{n}-\vec{n}^{\prime}).
 \end{eqnarray}
 From now on, $z$ is then implicitely replaced by its time-averaged
value. The signal on the bolometers is finally: 
\begin{eqnarray}
\mathcal{S}_{k} & = & \int\left|z_{k}(\vec{n})\right|^{2}\mathrm{d}\vec{n}\label{EqSignal}.
\end{eqnarray}

\section{Stokes parameter visibilities\label{stokes}}
Developping the signal on the bolometers in terms of the incoming electric fields easily shows autocorrelation terms for each channel as well as cross-correlation terms between all the possible pairs of channels:
\begin{eqnarray}
\mathcal{S}_k& =& \frac{1}{N_\mathrm{out}}\int\left\{   
\sum_{i=0}^{N_h-1} \left| \sum_{\eta=0}^{1}\epsilon_i^{\eta}(\vec{n})\exp\left(i\Phi^{\eta}_{k,i}\right)\right|^2 \right. \nonumber \\
&& \left. +2\mathrm{Re}\left[  
\sum_{i<j}\sum_{\eta_1,\eta_2}  \epsilon_i^{\eta_1}(\vec{n})\epsilon_j^{\eta_2\star}(\vec{n})   \exp\left(i(\Phi^{\eta_1}_{k,i}-\Phi^{\eta_2}_{k,j})\right)
\right]
\right\} \mathrm{d}\vec{n} \label{signal}.
\end{eqnarray}
The electric fields from different horns are related through Eq.~\ref{marchdiff} and introduce the Stokes parameters that are generally used to describe a polarized radiation:
\begin{eqnarray}
I=& \left< \left| E_\parallel\right|^2 \right> + \left< \left| E_\perp\right|^2 \right>, \\
Q=&  \left< \left| E_\parallel\right|^2 \right> - \left< \left| E_\perp\right|^2 \right>, \\
U=&  \left< E_\parallel  E^\star_\perp \right> + \left< E^\star_\parallel  E_\perp \right>  &= 2\mathrm{Re} \left< E_\parallel  E^\star_\perp \right>, \\
V=& i\left( \left< E_\parallel  E^\star_\perp \right> - \left< E^\star_\parallel  E_\perp \right>\right) &= -2\mathrm{Im} \left< E_\parallel  E^\star_\perp \right>.
\end{eqnarray}
The Stokes parameter visibilities are defined as ($S$ stands for $I$, $Q$, $U$ or $V$):
\begin{eqnarray}
\mathsf{V}_S(\vec{u}_\beta)=\int S(\vec{n})B_\mathrm{in}^2(\vec{n})\exp(2i\pi\vec{u}_\beta\cdot\vec{n})\mathrm{d}\vec{n}.
\end{eqnarray}
The phase-shift differences for a baseline $b$ formed by horns $i$ and $j$ measured in the channel $k$ are: 
\begin{eqnarray}
\Delta\Phi_{k,b}^{\parallel~\parallel}&=&\Phi_{k,i}^\parallel-\Phi_{k,j}^\parallel, \\
\Delta\Phi_{k,b}^{\perp\perp}&=&\Phi_{k,i}^\perp-\Phi_{k,j}^\perp, \\
\Delta\Phi_{k,b}^{\parallel\perp}&=&\Phi_{k,i}^\parallel-\Phi_{k,j}^\perp, \\
\Delta\Phi_{k,b}^{\perp\parallel}&=&\Phi_{k,i}^\perp-\Phi_{k,j}^\parallel.
\end{eqnarray}
Putting all these definitions into Eq.~\ref{signal} and after some calculations one finds that the signal on the bolometer $k$ can be expressed purely in terms of the Stokes parameter visibilities and the phase-shifting values  (the subscript $b$ stands for all the $N_h(N_h-1)/2$ available baselines and $n_k$ is the noise):
\begin{eqnarray}
\mathcal{S}_k&=&\vec{\Lambda}_k \cdot \vec{\mathsf{S}} +\sum_{b=0}^{N_h(N_h-1)/2-1} \vec{\Gamma}_{k,b} \cdot\vec{\mathcal{V}}_b +n_k \label{linearprob},
\end{eqnarray}
where the first term is the autocorrelations of all horns and the second one contains the cross-correlations, hence the interference patterns. We have used the following definitions:
{\small
\begin{eqnarray}
&&\vec{\Lambda}_k=\frac{1}{N_\mathrm{out}} \sum_{i=0}^{N_h-1}
\left(\begin{array}{c}
1 \\
0 \\
 \cos(\Phi_{k,i}^{\parallel}-\Phi_{k,i}^{\perp}) \\
 \sin(\Phi_{k,i}^{\parallel}-\Phi_{k,i}^{\perp})
\end{array}\right),
~
\vec{\mathsf{S}}^t=\left(
\begin{array}{c}
\int I(\vec{n})B^2(\vec{n})\mathrm{d}\vec{n} \\
\int Q(\vec{n})B^2(\vec{n})\mathrm{d}\vec{n} \\
\int U(\vec{n})B^2(\vec{n})\mathrm{d}\vec{n} \\
\int V(\vec{n})B^2(\vec{n})\mathrm{d}\vec{n} 
\end{array}\right), \\
&&\vec{\Gamma}_{k,b}=\frac{1}{N_\mathrm{out}}\left( \begin{array}{c}
\cos \Delta \Phi_{k,b}^{\parallel~\parallel} +\cos \Delta \Phi_{k,b}^{\perp \perp} \\
-(\sin \Delta \Phi_{k,b}^{\parallel~\parallel} +\sin \Delta \Phi_{k,b}^{\perp \perp}) \\
\cos \Delta \Phi_{k,b}^{\parallel~\parallel} -\cos \Delta \Phi_{k,b}^{\perp \perp} \\
-(\sin \Delta \Phi_{k,b}^{\parallel~\parallel} -\sin \Delta \Phi_{k,b}^{\perp \perp}) \\
\cos \Delta \Phi_{k,b}^{\parallel \perp} +\cos \Delta \Phi_{k,b}^{\perp \parallel}\\
-(\sin \Delta \Phi_{k,b}^{\parallel \perp} +\sin \Delta \Phi_{k,b}^{\perp \parallel})\\
-(\sin \Delta \Phi_{k,b}^{\parallel \perp} -\sin \Delta \Phi_{k,b}^{\perp \parallel})\\
-(\cos \Delta \Phi_{k,b}^{\parallel \perp} -\cos \Delta \Phi_{k,b}^{\perp \parallel)}
\end{array}
\right) 
,~
\vec{\mathcal{V}_b}^t=\left(\begin{array}{c}
\mathrm{Re}\left[{\vi}(\vec{u}_b)\right] \\
\mathrm{Im}\left[{\vi}(\vec{u}_b)\right] \\
\mathrm{Re}\left[{\vq}(\vec{u}_b)\right] \\
\mathrm{Im}\left[{\vq}(\vec{u}_b)\right] \\
\mathrm{Re}\left[{\vu}(\vec{u}_b)\right] \\
\mathrm{Im}\left[{\vu}(\vec{u}_b)\right] \\
\mathrm{Re}\left[{\vv}(\vec{u}_b)\right]\\
\mathrm{Im}\left[{\vv}(\vec{u}_b)\right]
\end{array} \right). \label{coeffs}
\end{eqnarray}}
All of this can be regrouped as a simple linear expression involving a vector with all the sky informations (Stokes parameter autocorrelations $\vec{\mathsf{S}}$ and all visibilities $\vec{\mathcal{V}_b}$) labelled $\vec{X}$ and another involving the phase-shifting informations ($\vec{\Lambda}_k$ and $\vec{\Gamma}_{k,b}$) labelled $\vec{A}_k$:
\begin{eqnarray}
\mathcal{S}_k&=& \vec{A}_k\cdot \vec{X}+n_k.
\end{eqnarray}
Finally, various measurements of the signal coming from the different channels and/or from different time samples with different phase-shifting configurations can be regrouped together by adding columns to $\vec{A}$ which then becomes a matrix $A$ and transforming the individual measurement $\mathcal{S}_k$ into a vector $\vec{\mathcal{S}}$:
\begin{eqnarray} \label{signalfinal}
\vec{\mathcal{S}}&=& A\cdot \vec{X}+\vec{n}.
\end{eqnarray}

\section{Reconstruction of the visibilities\label{reco}}
Once one has recorded enough data samples to invert the above linear problem (we will call such a period a {\em sequence} in the following), the solution is the usual one assuming that the measurements noise covariance matrix is $N=\left< \vec{n}\cdot \vec{n}^t\right>$:
\begin{eqnarray}
\vec{\hat{X}}=(A^t\cdot N^{-1}\cdot A)^{-1}\cdot A^{t}\cdot N^{-1}\cdot \vec{\mathcal{S}}, \label{solution}
\end{eqnarray}
with covariance matrix:
\begin{eqnarray}\label{covmat}
\mathcal{N}=\left< \left(\vec{\hat{X}}-\left< \vec{\hat{X}}\right>\right)\cdot\left(\vec{\hat{X}}-\left< \vec{\hat{X}}\right>\right)^t\right>=(A^t\cdot N^{-1}\cdot A)^{-1}.
\end{eqnarray}

\subsection{Regrouping equivalent baselines}
One sees that the dimension $N_u$ of the $\vec{X}$ vector of unknowns is rather large: $N_u=3+8\times N_b$ where $N_b=N_h(N_h-1)/2$ is the number of baselines formed by the input horn array. For a large horn array this number can become really large. A $10\times 10$ array has for instance $N_b=4950$ baselines and $N_u=39603$ unknowns. One needs at least as many data samples as unknowns (and in many cases  more than that) so this would involve manipulations of very large matrices. In fact as we said before, depending on the relative positions of the input horns, there may be a lot of {\em equivalent} baselines: different couples of horns separated by the same vector $\vec{u}_\beta$ hence measuring exactly the same visibilities. It is clearly advantageous to regroup these equivalent baselines together in order to reduce the dimension of the system. As we will see below there is a huge extra-advantage to do it this way in terms of signal-to-noise ratio if one chooses the phase-shifters angles wisely.

In the case where the input horn array is a square grid with size $N_\mathrm{side}=\sqrt{N_h}$, the number of different classes of equivalent baselines is $N_\neq=2N_{\mathrm{side}}(N_{\mathrm{side}}-1) =2(N_h-\sqrt{N_h})=180$ for a $10\times10$ horn array, hence reducing the number of unknowns to $1443$ which is a huge improvement. It is obvious that all equivalent baselines measure the same visibilities and can therefore be regrouped together in the linear problem leading to the same solution as considering the equivalent baselines separately. One just has to reorder the terms in Eq.~\ref{linearprob} as first a sum over all different baselines $\beta$ and then a sum over each of the baselines  $b_\beta$ equivalent to $\beta$ coming on the output line $k$:
\begin{eqnarray}
\mathcal{S}_k&=&\vec{\Lambda}_k \cdot \vec{\mathsf{S}} +\sum_{\beta=0}^{N_\neq-1} \vec{\Gamma}_{k,\beta} \cdot\vec{\mathcal{V}}_{\beta}+n_k \label{newlinearprob},
\end{eqnarray}
changing the $\vec{\Gamma}$ vector to:
\begin{eqnarray}
\vec{\Gamma}_{k,\beta}=\frac{1}{N_\mathrm{out}}\sum_{b_\beta=0}^{N_\mathrm{eq}(\beta)-1} \left( \begin{array}{c}
 \cos \Delta \Phi_{k,b_\beta}^{\parallel~\parallel} +\cos \Delta \Phi_{k,b_\beta}^{\perp \perp} \\
 -(\sin \Delta \Phi_{k,b_\beta}^{\parallel~\parallel} +\sin \Delta \Phi_{k,b_\beta}^{\perp \perp}) \\
 \cos \Delta \Phi_{k,b_\beta}^{\parallel~\parallel} -\cos \Delta \Phi_{k,b_\beta}^{\perp \perp} \\
 -(\sin \Delta \Phi_{k,b_\beta}^{\parallel~\parallel} -\sin \Delta \Phi_{k,b_\beta}^{\perp \perp}) \\
 \cos \Delta \Phi_{k,b_\beta}^{\parallel \perp} +\cos \Delta \Phi_{k,b_\beta}^{\perp \parallel}\\
 -(\sin \Delta \Phi_{k,b_\beta}^{\parallel \perp} +\sin \Delta \Phi_{k,b_\beta}^{\perp \parallel})\\
 -(\sin \Delta \Phi_{k,b_\beta}^{\parallel \perp} -\sin \Delta \Phi_{k,b_\beta}^{\perp \parallel})\\
 -(\cos \Delta \Phi_{k,b_\beta}^{\parallel \perp} -\cos \Delta \Phi_{k,b_\beta}^{\perp \parallel)}
\end{array}
\right). \label{newcoeffs}
\end{eqnarray}
Let's recall that each column of the matrix $A$ corresponds to phase-shifters configurations encoded in $\vec{\Gamma}_{k,\beta}$ (for all different baselines $\beta$) and $\vec{\Lambda}_k$.

\subsection{Coherent summation of equivalent baselines}

We shall now investigate the noise covariance matrix for the reconstructed
visibilities and how one can possibly optimize it. We will assume
for simplicity that the noise is stationary and uncorrelated from
one data sample to another where we call data sample the output of
one of the $N_\mathrm{out}$ output channel during one of $N_{t}$ time samples.
Therefore there are $N_{d}=N_\mathrm{out}\times N_{t}$ data samples. If the
photon noise corresponding to one horn on one detector measured during
one time sample is $\sigma_{0}$, then the noise covariance matrix
of the measured data samples is: 
\begin{equation}
N=\frac{\sigma_{0}^{2}N_{h}}{N_\mathrm{out}}\times\bbbone ,
\end{equation}
 where $\bbbone$ is the $N_{d}\times N_{d}$ identity matrix. Assuming
that the time variation of the $A$ matrix can be neglected%
\footnote{Time variation of the $A$ matrix is of course a source of systematic
errors and must be studied as such.%
}, It can be trivially extended to a $N_{d}\times N_{h}$ matrix $A_{k\, t_{i},\beta}=A_{k,\beta}$. \\
In terms of this extended $A$ matrix, the visibilities covariance
matrix (see Eq.~\ref{covmat}) writes: 
\begin{eqnarray}
\mathcal{N}=\frac{\sigma_{0}^{2}N_{h}}{N_\mathrm{out}}\times\left(A^{t}\cdot A\right)^{-1}.
\end{eqnarray}

We have regrouped all equivalent baselines together in $A$, each
of its elements is therefore the sum on $N_{\mathrm{eq}}$ sines and
cosines of the phase-shifting angles (as expressed in Eq.~\ref{newcoeffs}).
We will assume here that the angles are chosen randomly and uniformly
from a set of possible values between 0 and $2\pi$. Now there are
two possibilities depending on the choice for the phase-shifting angles
for all baselines equivalent to a given one: they can all be different
or they can all be equal. We refer to this choice as {\em incoherent
or coherent} summation of equivalent baselines: 

\begin{itemize}
\item {\em Incoherent summation of equivalent baselines:} each of the
sum of the two sine/cosine functions of the uniformly distributed
angles has zero average and a variance 1. Each element of $\vec{\Gamma}_{k,\beta}$
is the sum of $N_{\mathrm{eq}}$ of these and the {\em Central Limit
Theorem} states that it will have zero average and a variance $\left(\frac{1}{N_\mathrm{out}}\right)^{2}N_{\mathrm{eq}}(\beta)$. 
\item {\em Coherent summation of equivalent baselines:} then each elements
of $\vec{\Gamma}_{k,\beta}$ is $\frac{1}{N_\mathrm{out}}N_{\mathrm{eq}}(\beta)$
times the same angle contribution with variance 1. The matrix elements
ends up having a variance $\left(\frac{1}{N_\mathrm{out}}\right)^{2}N_{\mathrm{eq}}^{2}(\beta)$. 
\end{itemize}
Coming back to $A^{t}\cdot A$, the multiplication by the transpose
will add together all the $N_{d}$ different data samples. The off-diagonal
elements will cancel out to zero because the angles are uncorrelated
from one channel to another. The diagonal elements will however average
to the variance of the elements in $A$ multiplied by $N_{d}$. So
finally, depending on the choice between incoherent or coherent summation
of equivalent baselines, the visibility covariance matrix will scale
in a different manner: \begin{eqnarray} \label{eqsummation}
\mathcal{N}=\left\{ \begin{array}{lll}
\frac{\sigma_{0}^{2}N_{h}}{N_\mathrm{out}}\frac{1}{N_{d}}\frac{N_\mathrm{out}^{2}}{N_{\mathrm{eq}}(\beta)} & =\frac{\sigma_{0}^{2}N_{h}}{N_{t}}\frac{1}{N_{\mathrm{eq}}(\beta)} & ~~\mathrm{for~incoherent~summation},~\\
~\\
\frac{\sigma_{0}^{2}N_{h}}{N_\mathrm{out}}\frac{1}{N_{d}}\frac{N_\mathrm{out}^{2}}{N_{\mathrm{eq}}^{2}(\beta)} & =\frac{\sigma_{0}^{2}N_{h}}{N_{t}}\frac{1}{N_{\mathrm{eq}}^{2}(\beta)} & ~~\mathrm{for~coherent~summation}.\end{array}\right.\end{eqnarray}
 The latter scaling is clearly more advantageous and optimises the
reconstruction of the visibilities. In fact this result is quite obvious: if the phase-shifting angles for equivalent baselines are all different,
the coefficients of the linear problem that one wants to invert will
always be smaller than if the summation of equivalent baselines is
performed coherently. 
The signal to noise ratio on the visibilities
will therefore be optimal if one maximises the coefficients, which
is obtained by choosing the coherent summation.

\subsection{Comparison with classical interferometers and imagers}
The variance on the visibilities obtained above in the case of a coherent summation of equivalent baselines can be rewritten:
\begin{equation}
\sigma_{\mathsf{V}(\beta)}^2=\frac{N_{h}}{N_{\mathrm{eq}}(\beta)}\frac{\sigma_{0}^{2}}{N_{t}N_{\mathrm{eq}}(\beta)},
\end{equation}
that can be compared\footnote{The notations are different : $t_\mathrm{vis}$ in~[\cite{hobson}] has to be replaced by our $N_t$, their $n_\mathrm{vis}$ is the number of equivalent baselines $N_\mathrm{eq}$. In our article $\sigma_0$ corresponds to $s\Omega_s$ in their article as a noise equivalent power $\mathrm{NEP}$ has to be replaced by $\mathrm{NET}\times \Omega$ when talking about noises in temperature unit rather than in power unit.} to formula (28) in~[\cite{hobson}]  which is the equivalent for heterodyne interferometry $\sigma_{0}^{2}/N_{t}N_{\mathrm{eq}}(\beta)$. We see that the only difference introduced by bolometric interferometry is the factor $N_h/N_\mathrm{eq}(\beta)$. In average, the number of equivalent baselines is $\left<N_{\mathrm{eq}}\right> = (N_{h}(N_{h}-1)/2)/N_\neq\simeq N_{h}/4$, but is much larger for small baselines. The design of the instrument has to be such that the "interesting" baselines are very redundant leading to a $N_h/N_\mathrm{eq}$ closer to one. The resulting expression of the variance on visibilities for bolometric interferometry therefore only differs by this slightly larger than one factor with respect to heterodyne interferometry. The important point is that the value of $\sigma_0$ for bolometric interferometry is typical of a bolometer (photon noise dominated) hence smaller than what can be achieved with HEMT amplifiers in a heterodyne interferometer.  
 
 This result can be summarized as follows: a bolometric interferometer using coherent summation of equivalent baselines can achieve the sensitivity that would be obtained with an heterodyne interferometer with the noise of a bolometric instrument (and without the complexity issues related to the large number of channels). Such an instrument would therefore be competitive with an imager that would have the same number of bolometers as we have input channels in our bolometric interferometer. A detailed study of the comparison between a bolometric interferometer and an imager from the sensitivity point of view is in preparation~[\cite{charlassierPrep}].
 On the
opposite, if the equivalent baselines are summed incoherently, it is obvious that the sensitivity would be very poor due to the absence of the $1/N_\mathrm{eq}$ additional factor.

The next section shows how it is possible to choose the phase-shifting
sequences in such a way that the prescription of {\em coherent summation
of equivalent baselines} is enforced.

\section{Choice of the optimal phase-sequences\label{phases}}

One wants the phase-shifting scheme to be such that {\em equivalent
baselines} have exactly the same sequence but that {\em different
baselines} have different phase-shifts so that they can be disentangled
by the linear inversion corresponding to Eq.~\ref{newlinearprob}.
Now let's see how to comply with this constraint of having {\em
equivalent} baselines correspond to identical phase differences.
An important remark is that, as can be seen in Eq.~\ref{phaseshift},
the phase-shift have two different origins: the phase-shifters themselves
on the one hand whose angles can be chosen to follow a given sequence
and are the same for all output channels and on the other hand, the
phase-shifts coming from the beam combiner. Each input will be labelled
by the horn number $0\le i\le N_{h}-1$ and the polarization direction
$\eta$. Each output is labelled by its number $1\le k\le N_\mathrm{out}$
. The phase-shift differences are therefore 
\begin{equation}
\Delta\Phi_{k,i\eta,j\eta'}=\underbrace{(\phi_{i\eta}-\phi_{j\eta'})}_{\mathrm{phase-shifters}}+\underbrace{(\psi_{k,i\eta}-\psi_{k,j\eta'})}_{\mathrm{beam~combiner}}.
\end{equation}

\begin{figure}[!t]
\centering\resizebox{4.5cm}{!}{\centering{
\includegraphics[angle=-90]{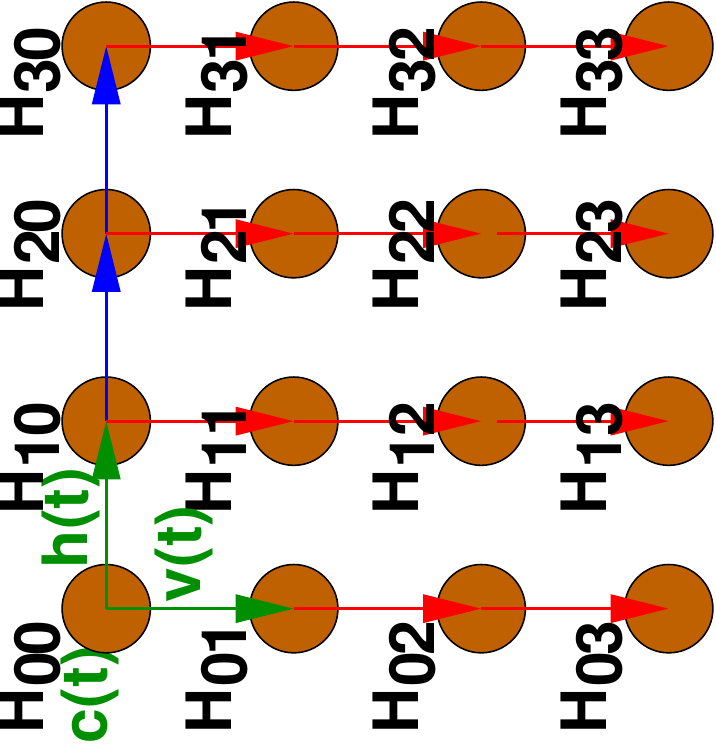}}}
\caption{\small Choosing all the phase sequences from that of one horn and two phase sequence differences (represented in green).}
\label{figseq}
\end{figure}

\paragraph*{a) Phase-shifter phase differences (unpolarized case):\\}

We assume that the horns are placed on a square array with size $N_{\mathrm{side}}=\sqrt{N_{h}}$
as in Fig.~\ref{figseq}. In this case, the position of all horns
can be parametrized, in units of the minimum horn separation, as a
vector $\vec{d}_{i}=\left(\begin{array}{c}
l_{i}\\
m_{i}\end{array}\right)$$ $ where $l_{i}$ and $m_{i}$ are integers running from 0 to $N_{\mathrm{side}}-1$
such that $i=l_{i}+N_{\mathrm{side}}m_{i}$. In this case, we have seen that there are
$N_\ne=2(N_h-\sqrt{N_h})$ different classes of equivalent
baselines labelled $\vec{u}_{\beta}$. Forgetting about polarization,
the phase sequences can be constructed from a vector of two independent
random phase sequences $h(t)$ and $v(t)$ which separate the horizontal
and vertical directions in the horn array: 
\begin{equation}\label{eqphase0}
\phi_{i}(t)=\vec{d}_{i}\cdot\vec{s}(t)\mbox{ where }\vec{s}(t)=\left(\begin{array}{c}
h(t)\\
v(t)\end{array}\right).
\end{equation}
The phase shift difference associated with the baseline between horns
$i$ and $j$ is 
\begin{equation}\label{eqphase}
\phi_{i}-\phi_{j}=(\vec{d}_{i}-\vec{d}_{j})\cdot\vec{s}(t),
\end{equation}
and it is clear that the phase shift difference sequences will be the same for
all baselines such that $\vec{d}_{i}-\vec{d}_{j}=\vec{u}_{\beta}$,
where $\beta$ is one of the classes of equivalent baselines. Because
the two random sequences $h(t)$ and $v(t)$ have been chosen independent,
the phase sequences associated with two different baselines classes
$\beta\neq\beta'$ will be different.

\paragraph*{b) Separating polarizations:\\}

Looking at formula (\ref{coeffs}), it is clear that one will not
be able to separate $\vi$ and $\vq$ visibilities, unless one uses\textbf{
}two independent vectors of sequences $\vec{s}_{||}(t)\neq\vec{s}_{\bot}(t)$.
However, in this case $\vu$ and $\vv$ are not measured with maximum
accuracy because the phase shift differences \[
\phi_{i\,||}-\phi_{j\,\bot}=\vec{s}_{||}(t)\cdot\vec{d}_{i}-\vec{s}_{\bot}(t)\cdot\vec{d}_{j},\] are not equal for two different but equivalent baselines, so that they do not add coherently. One is therefore led
to use alternately two measuring modes:

\begin{enumerate}
\item One mode where $\vec{s}_{||}(t)\neq\vec{s}_{\bot}(t)$, where phase
shifts differences read:\[
\phi_{i\eta}-\phi_{j\eta}=\vec{s}_{\eta}(t)\cdot(\vec{d}_{i}-\vec{d}_{j}).\]
In this mode, $\vi$ and $\vq$ are measured with maximum accuracy
(noise reduction $\propto N_{\mathrm{eq}}^{2}$), but $\vu$ and $\vv$
are only measured with noise a reduction $\propto N_{\mathrm{eq}}$.
\item One mode with $\vec{s}_{||}(t)=\vec{s}_{\bot}(t)=\vec{s}(t)$. Then
however, one cannot measure $\vv$ because \[
\phi_{i\,||}-\phi_{j\,\bot}=\vec{s}(t)\cdot(\vec{d}_{i}-\vec{d}_{j})=\phi_{i\,\bot}-\phi_{j\,||},\]
therefore one must introduce two more sequences $c_{||}(t)\neq c_{\bot}(t)$
(one of them may be zero) independent from one another and from $\vec{s}(t)$,
such that $\phi_{i\eta}=\vec{s}(t)\cdot\vec{d}_{i}+c_{\eta}(t)$.
Then:\textbf{\begin{eqnarray*}
\phi_{i\,||}-\phi_{j\,\bot} & = & \vec{s}(t)\cdot(\vec{u}_{i}-\vec{u}_{j})+c_{||}(t)-c_{\bot}(t)\\
 & \mathrm{whereas}\\
\phi_{i\,\bot}-\phi_{j\,||} & = & \vec{s}(t)\cdot(\vec{u}_{i}-\vec{u}_{j})+c_{\bot}(t)-c_{||}(t)\\
 & \mathrm{but}\\
\phi_{i\,||}-\phi_{j\,||} & = & \phi_{i\,\bot}-\phi_{j\,\bot}=\vec{s}(t)\cdot(\vec{u}_{i}-\vec{u}_{j}),\end{eqnarray*}
}which means that $\vi$, $\vu$ and $\vv$ are measured with maximum
accuracy (noise reduction $\propto N_{\mathrm{eq}}^{2}$), but $\vq$
is not measured at all.
\end{enumerate}

We mentioned before that we need all phase-shifters to be identical:
they all have to be able to produce the same $n_{\phi}$ phase-shifts
(let us call this ensemble $\Psi$) but in different order. If $h_{\eta}(t)$,
$v_{\eta}(t)$ and $c_{\eta}(t)$ are sequences of elements belonging
to $\Psi$, then the phase for any horn also has to belong to $\Psi$,
meaning that $\phi_{i\eta}(t)=l_i\, h_{\eta}(t)+m_i\, v_{\eta}(t)+c_{\eta}(t)$
has to belong to $\Psi$. As shown in Appendix~\ref{proof}, this
requires to choose the $n_{\phi}$ values of the phase-shifts regularly
spaced between 0 and $2\pi$ as: 
\begin{equation}
\phi_{n}=n\frac{2\pi}{n_{\phi}}~~~~~~~~~~~~~~~~(n=0,\dots ,n_{\phi}-1).
\end{equation}
The elementary sequences $h_{\eta}(t)$, $v_{\eta}(t)$ and $c_{\eta}(t)$
are uniform random sample of $N_{s}$ values taken among the $n_{\phi}$
elements of $\Psi$. They must be chosen independent from one another
to make sure that unequivalent baselines do not share the same sequence
of phase differences.

\paragraph*{c) Beam combiner phase difference:\\}

As was said before there are two main designs for the beam combiner: Butler combiner~[\cite{thesebutler}] or quasi optical combiner~[\cite{MBI}].
Without going into details, let us say that identical phase shifts
for equivalent baselines are naturally obtained for the quasi optical
combiner, and are achieved through an adequate wiring for the Butler
combiner.

\paragraph*{d) Summary and expected accuracy:\\}
Finally, in order to recover the visibilities keeping to the {}``coherent
summation of equivalent baselines'' criterion, one only has to build
phase sequences that successively follow modes 1 and 2 on an equal
footing, build the corresponding $A$ matrix and solve the system.
There is a price to pay: during the first half sequence, $\vi$ and
$\vq$ are measured with optimal accuracly but $\vu$ and $\vv$ are
not, during the second half sequence, $\vi$, $\vu,$ and $\vv$ are
measured with optimal accuracy but $\vq$ is not measured at all.
We therefore expect the sensitivity on $ $$\vq$, $\vu$ and $\vv$
to be down by roughly a factor of $\sqrt{2}$ with respect to the
sensitivity on $\vi$, although the sensitivity on $\vu$ and $\vv$
will be slightly better constrained than on $\vq$.

\section{Monte-Carlo simulations\label{mc}}
We have investigated what was discussed above using Monte-Carlo simulations. 
There are three approaches that have to be compared for the reconstruction of the Stokes parameter visibilities.
\begin{itemize}
\item Considering all baselines independantely without regrouping the equivalent ones. We expect this
method to have error bars scaling as $1/\sqrt{N_\mathrm{eq}}$. The system to solve is large in that case.
\item Regrouping the equivalent baselines together but without any choice for the phase-shifts so that they don't add in a coherent way. We expect this method to be exactly equivalent to the previous one but with a reduced size of the matrices.
\item Following the strategy to regroup equivalent baselines and choose the phases so that they are coherently added. We expect the error bars to scale as $1/N_\mathrm{eq}$ and therefore be the most efficient.
\end{itemize}
In each case, we have simulated random visibilities with $\vq$, $\vu$ and $\vv$ a hundred times lower than $\vi$ as expected from the CMB and calculated the signal expected on the bolometers using the phase-shift values for the three above strategies. We then added Gaussian noise with a variance $\sigma^2_\mathrm{MC}=\sigma_0^2 N_h/N_t$ to the bolometer signal. In each case we have performed a large number of noise and phase-shift sequence realisations. For each realisation, we have stored the reconstructed and input visibilities and analysed the residuals distributions. We have investigated the three above strategies and also the behaviour of the third one (coherent summation of equivalent baselines) with respect to the two free parameters: the length of the phase-shift sequence before inverting the linear problem and the number of different phase-shift angles (regularly spaced between 0 and $2\pi$ as shown in Appendix~\ref{proof}). 
\begin{figure*}[!ht]
\centering\resizebox{\hsize}{!}{\centering{
\includegraphics[angle=0]{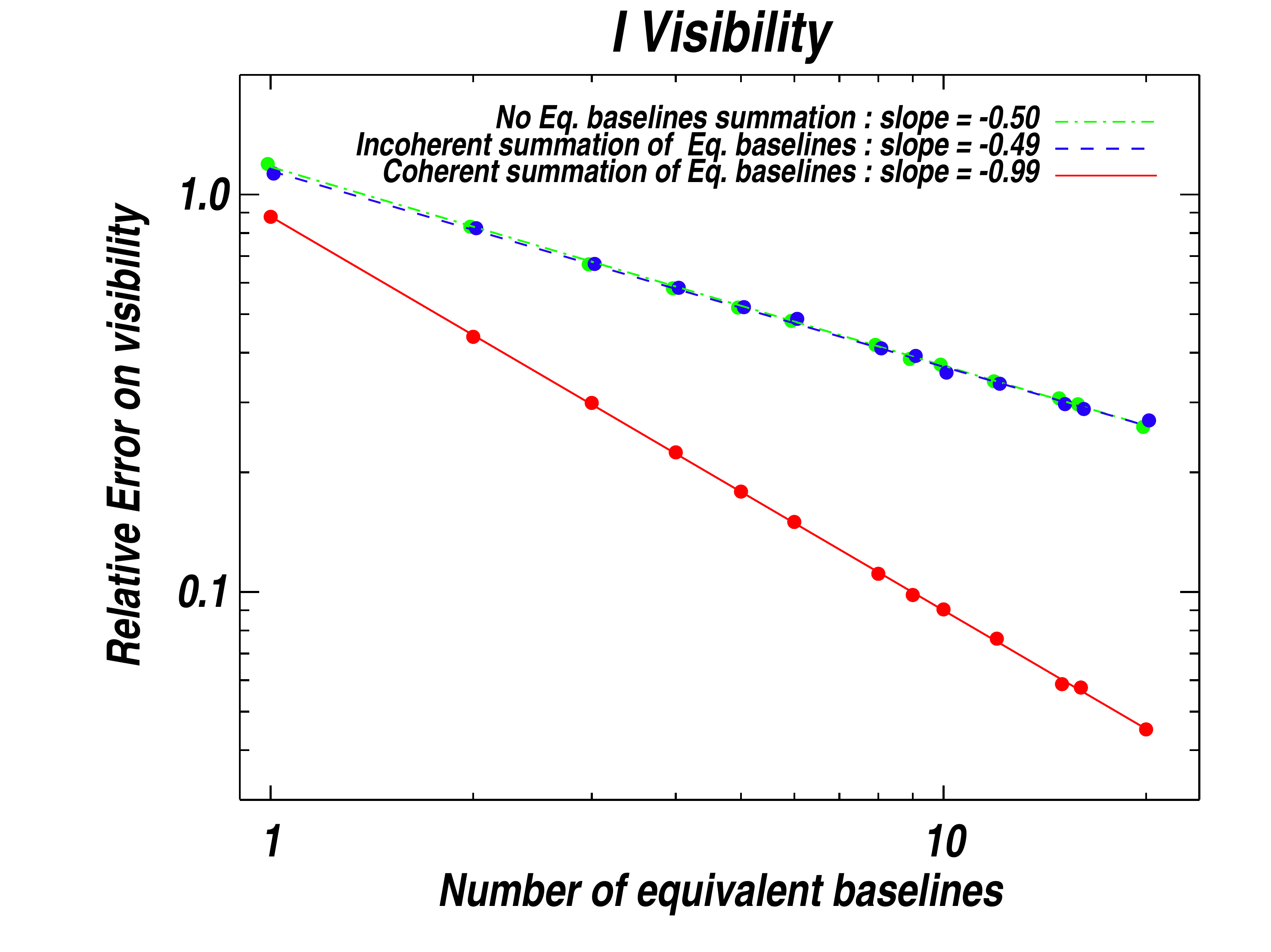}\includegraphics[angle=0]{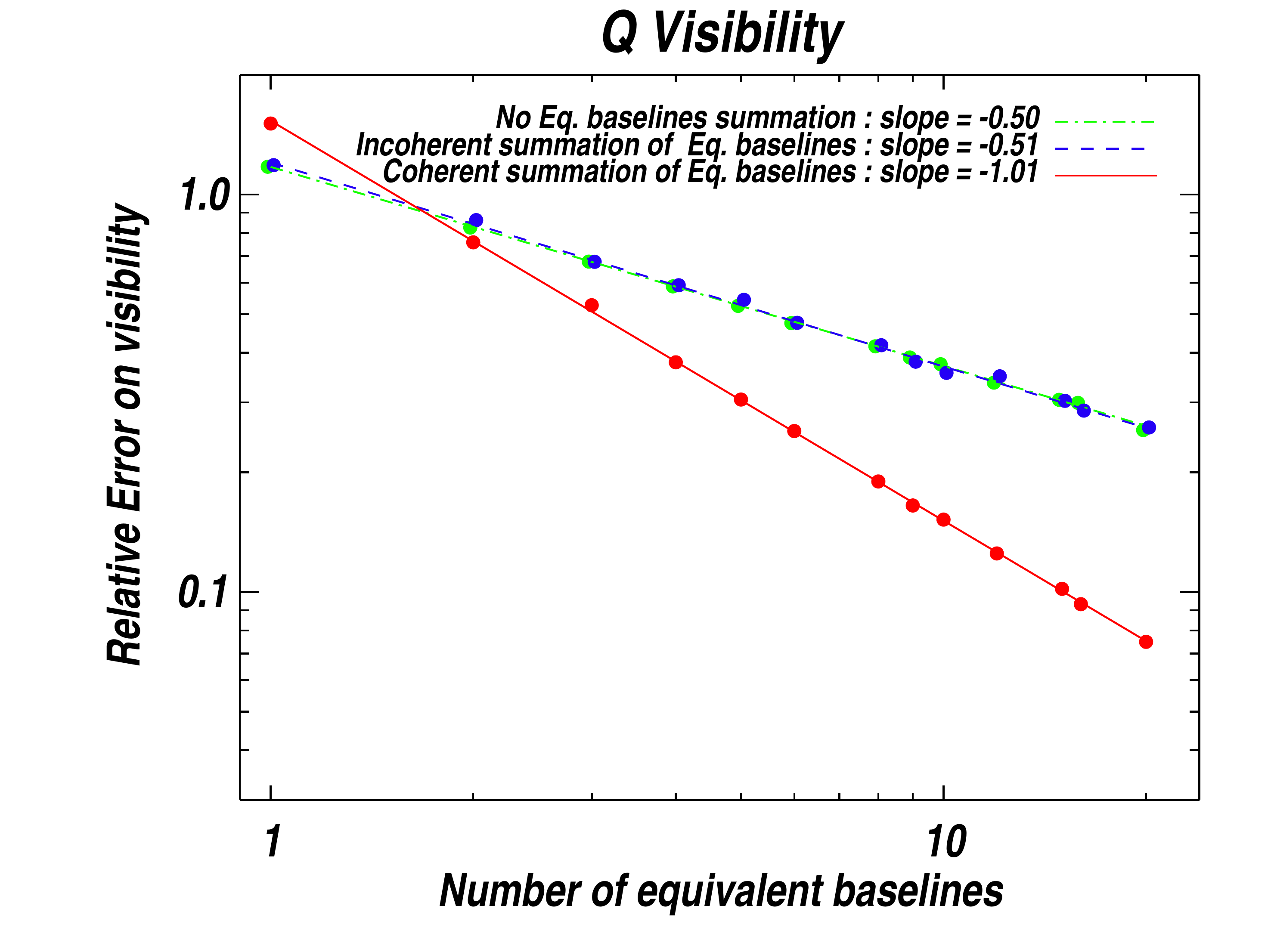}}}
\caption{\small Relative RMS  on visibility residuals for $\vi$ (left) and $\vq$ (right). The behaviour for $\vu$ and $\vv$ would be exactly the same as for $\vq$. The RMS on the residuals has been divided by the expected $\sigma_0\times\sqrt{\frac{N_h}{N_t}}$ scaling for each strategy to exhibit only the dependence with the number of equivalent baselines. The data points were fitted with linear slopes in Log-Log to measure the power of the scaling. One sees that the strategy where equivalent baselines are summed in a coherent way leads to a much better scaling $\propto\frac{1}{N_\mathrm{eq}}$ than the other strategies that both scale as $\propto\frac{1}{\sqrt{N_\mathrm{eq}}}$.}
\label{figneq}
\end{figure*}

\subsection{Scaling with the number of equivalent baselines}
We show in Fig.~\ref{figneq} the scaling of the RMS residuals on the visibilities as a function to the number of equivalent baselines. We have divided the RMS by $\sigma_0\sqrt{N_h/N_t}$ in order to isolate the effects that are specific to bolometric interferometry and depend on the way equivalent baselines are summed (see Eq.~\ref{eqsummation}).
We see that as expected the scaling is $\propto 1/N_\mathrm{eq}$ if one solves the problem by maximizing the signal to noise ratio using our coherent summation of equivalent baselines. The poor $1/\sqrt{N_\mathrm{eq}}$ scaling is also observed when all baselines are considered separately or when the phase-shift angles are not choosen optimally.

\subsection{Scaling with the number of samples and number of different phases}
\begin{figure*}[!ht]
\centering\resizebox{\hsize}{!}{\centering{
\includegraphics[angle=0]{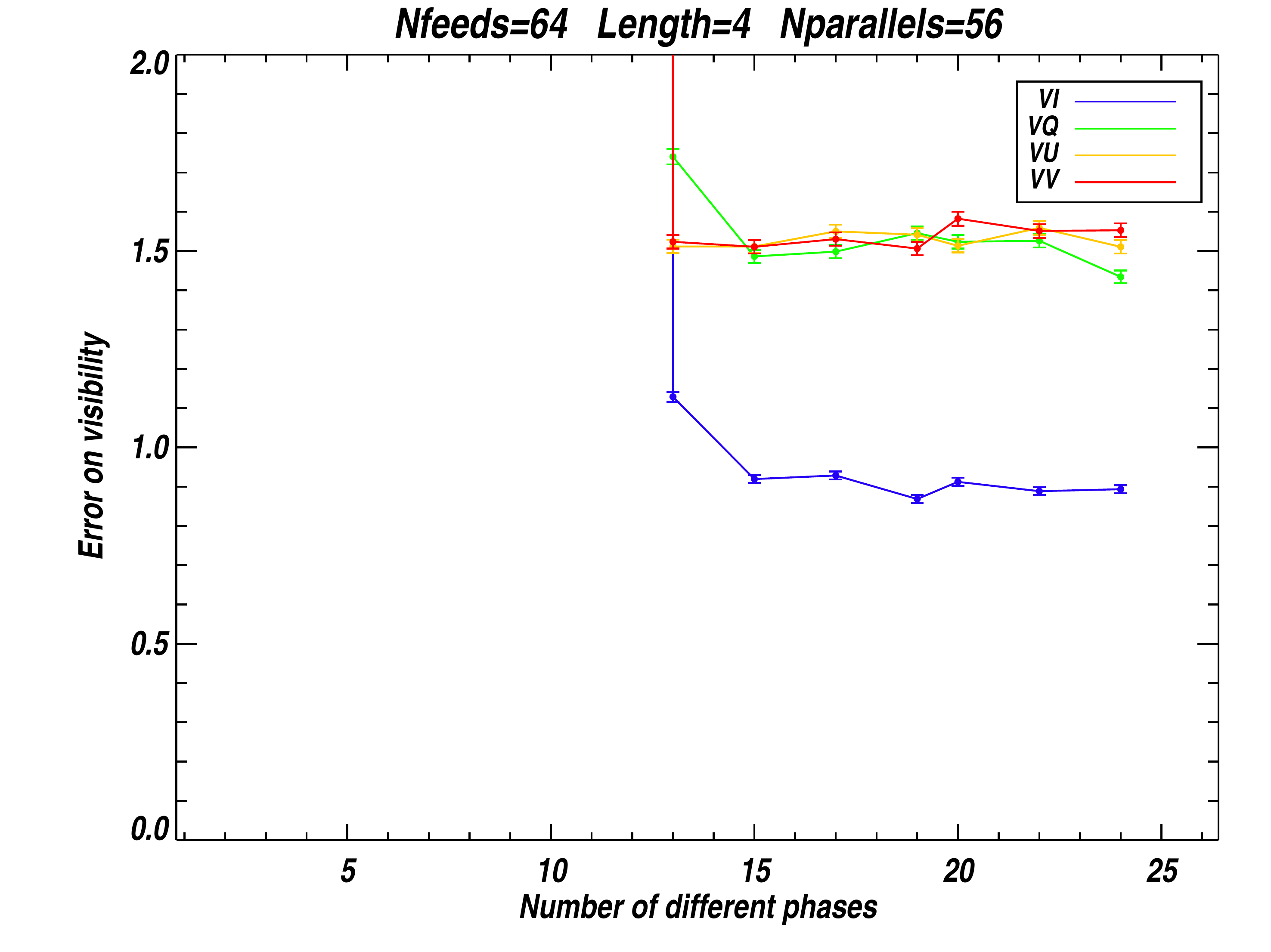}\includegraphics[angle=0]{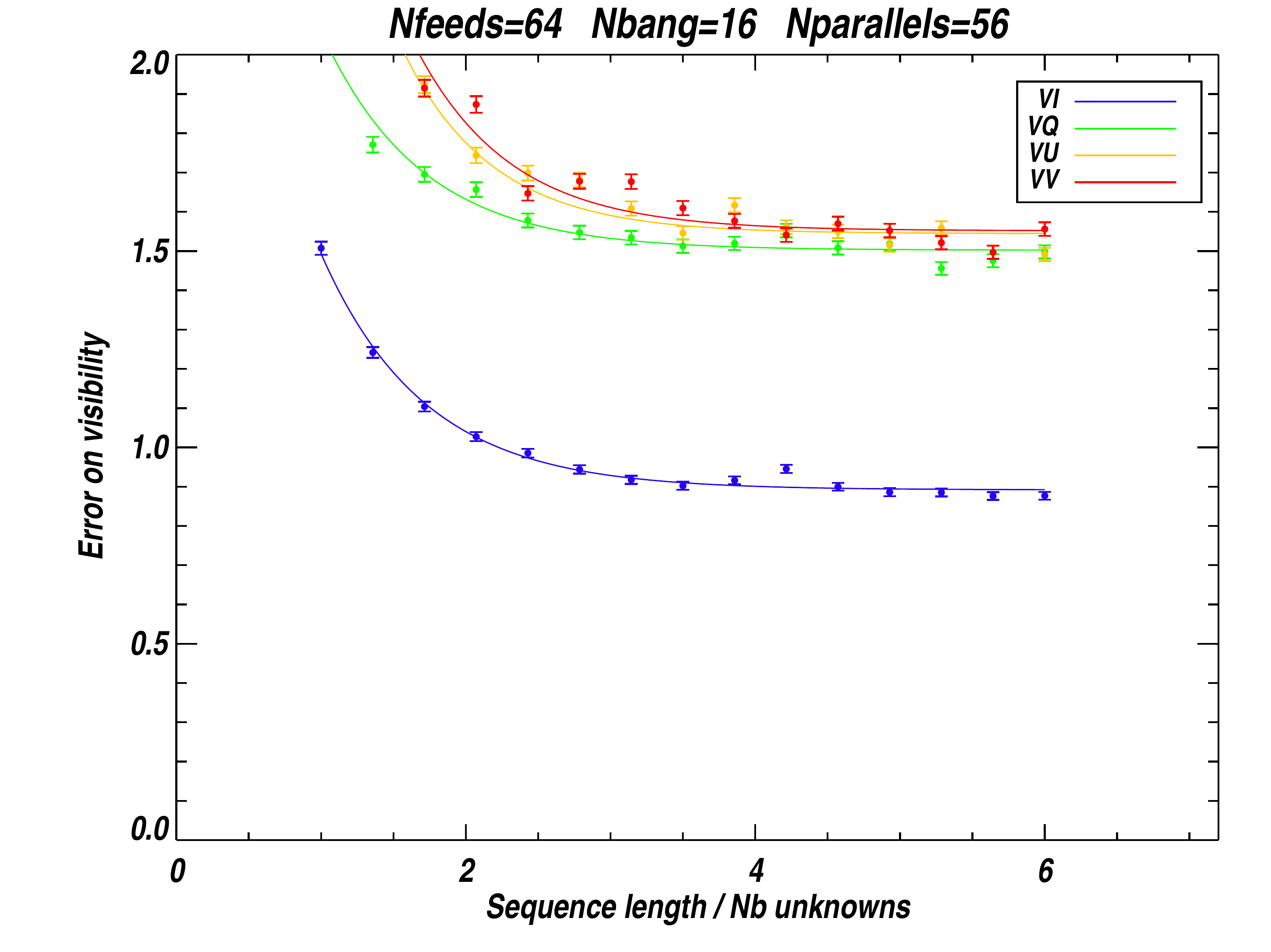}}}
\caption{\small Scaling of the RMS residuals (divided by  $\frac{\sigma_0}{N_\mathrm{eq}}\sqrt{\frac{N_h}{N_t}} $) on the Stokes parameter visibilities with respect to the number of different phases achieved by the phase-shifters and the number of different phase configurations used for the analysis (length of the sequence). One sees on the left that for a number of horns of $64$ one needs at least 12 or 13 different angles to be able to solve the linear problem. It is clear from the plot on the right that the longer the sequence, the better the residuals, but a plateau is rapidly reached when the number of samples is around 4 times the number of unknowns in the linear problem. One can also see the factor $\simeq\sqrt{2}$ between the accuracy on the intensity and the polarized Stokes parameters due to the two-steps phase-shifting scheme that we have to perform to be able to reconstruct them all.}
\label{scaling}
\end{figure*}

Let's now concentrate on the optimized strategy described above: coherent summation of equivalent baselines.
We show in Fig.~\ref{scaling} the scaling of the RMS residuals on the visibilities with respect to the length of the sequence and the number of different phases achieved by the phase-shifters (as shown in Appendix~\ref{proof}, these have to be regularly spaced between 0 and $2\pi$). The RMS values have been divided by  $\frac{\sigma_0}{N_\mathrm{eq}}\sqrt{\frac{N_h}{N_t}} $.

 One observes (Fig.~\ref{scaling} left) that the linear problem is singular when the number of different phases is not sufficient. Varying the number of horns in the array led us to derive the general scaling $\simeq 2\sqrt{N_h}$ for the minimum number of phases. Increasing the number of possible angles does not improve the residuals. Concerning the length of the sequence (Fig.~\ref{scaling} right), one observes that when it is slightly larger than the number of unknows ($N_u=3+8\times N_\neq$ where $N_\neq$ is the number of different baselines, $N_\neq=2(N_h-\sqrt{N_h})$ for a square array) then the reconstruction of the visibilities is not optimal due to the lack of constraints. Optimality is progressively reached when integrating a larger number of samples before inverting the problem. A reasonnably optimal result is obtained when $N_d\simeq 4\times N_u$. The expected $\simeq\sqrt{2}$ difference between the accuracy on $\vi$ and that on $\vq$, $\vu$ and $\vv$ (due to the fact that we have have to perform two successive phase-shifting schemes in order to measure all three polarized visibilities) is also confirmed by the simulation.
 
\section{How to proceed with a realistic instrument ?\label{syste}} 
When dealing with a realistic instrument one has to account for systematic errors and uncertainty to choose the precise data analysis strategy. We do not want to address the wide topic of systematic effects with bolometric interferometry  in this article (we refer the interested reader to~[\cite{bunn}] where systematic issues for interferometry are treated in a general way) but just want to stress one point that is specific to the method we propose here, related to intercalibration of the bolometers in the detector array. 

Inverting the linear problem in Eq.~\ref{newlinearprob} is nothing but expressing the Stokes parameter visibilities as linear combinations of the $N_d$ signal measurements performed with different phase-shifting configurations. These measurements can be those of the $N_\mathrm{out}$ bolometers each in $N_t$ time samples. This is where intercalibration issues have to be considered. Linear combinations of signals measured by different bolometers are extremely sensitive to errors in intercalibration and will induce leakage of intensity into the polarized Stokes parameters if it is not controlled up to an exquisite accuracy. So we claim that combining different bolometers in the reconstruction of the visibilities in a bolometric interferometer such as the one we describe here is not a wise choice unless the bolometers array is very well intercalibrated (through precise flat-fielding). The solution we propose is to treat all the bolometers independantly, inverting the linear problem separately for each of them. This requires a lot of time samples for the phase-shift sequences but is safer from the point of view of systematics. As a realistic example, for a $10\times 10$ elements square input array, the number of different baselines is 180 and the number of unknowns is 1443. An optimal reconstruction of the visibilities can therefore be achieved with $\sim$6000 time samples. The duration of the time samples is driven by both the time constant of the bolometers (very short with TES) and the speed achieved by the phase-shifter to switch from one phase to the other. A reasonnable duration for the time samples is about 10 msec which would correspond to sequences lasting about one minute. It is likely that the cryogenic system of such a bolometric interferometer would ensure a stable bath on the minute time scale so that the knee frequency of the bolometric signal would be smaller than 1~min$^{-1}$. In such a case, the noise can be considered as white (diagonal covariance matrix) during each sequence and the inversion gets easily tractable even with ~6000 samples vectors. We are currently performing fully realistic simulations including systematic effects, the results will be presented in a future publication.

\section*{Conclusions}
We have investigated the way to reconstruct the Stokes parameter visibilities from a bolometric interferometer. It turns out that all three complex Stokes parameter visibilities can be reconstructed with an accuracy that scales as the inverse of the number of equivalent baselines if one follows a simple prescription: all equivalent baselines have to be factorized together in a coherent way, meaning that the phase-shift differences have to be equal for equivalent baselines. We have proposed a simple way to construct such phase-shift sequences and tested it on a Monte-Carlo simulation. The simulation confirms that the scaling of the errors on the visibilitites is $\propto \sqrt{N_h}/N_\mathrm{eq}$ if one follows our prescription but $\sqrt{N_h/N_\mathrm{eq}}$ otherwise. 

The main conclusion of this article is therefore that a bolometric interferometer is competitive with an imager having the same number of horns (instrumental noise on the power spectrum $\propto 1/N_h$) but only 
with an appropriate choice of the phase-shift sequences (coherent summation of equivalent baselines).

We also discussed the data analysis strategy and proposed a solution to the possible cross-calibration issues between the different bolometers. Even though one has simultaneously $N_\mathrm{out}$ measurements of the signal with different phase configuration, it might be preferable not to combine these measurements but to reconstruct the visibilities on each bolometer separately and combine the visibilities afterwards. Such a strategy would increase the length of the phase-shifting sequences, but in a reasonable (and tractable) way thanks to the intrinsic shortness of our proposed phase-shifting scheme.

\begin{acknowledgements}
      The authors are grateful to the whole BRAIN collaboration for fruitful discussions.
\end{acknowledgements}

\begin{appendix}
\section{Proof of the necessity of having regularly spaced phase-shift values\label{proof}}
When we use the phase-shift configurations of Eq.~\ref{eqphase0}, the antenna with coordinates $(i,j)$ will be phase-shifted by:
\begin{equation}\label{eqapp}
\phi_{i,j}(t)=ih(t)+jv(t)+c(t).
\end{equation}
In practice we are only able to construct a limited number of different phase-shifters, and the phase-shift sequences $h(t)$, $v(t)$ and $c(t)$ will be independent random sequences of phase-shifts taken from {\bf the same set $\Phi$ of $n$ phase-shifts $\phi_p$}. For all phase-shifts in Eq.~\ref{eqapp} to belong to $\Phi$, it is necessary that  $l\times \phi_p$ (modulo $2\pi$) also belongs to $\Phi$. Let us write the smallest non-zero element of $\Phi$ as:
\begin{equation}
\phi_\mathrm{min}=\frac{2\pi}{n+\epsilon}~,     ~~~~~~ n\in \mathrm{I\!N},~~~~~0\leq\epsilon< 1.
\end{equation}
$(n+1)\phi_\mathrm{min}$ (modulo $2\pi$) should also belong to $\Phi$, but
\begin{equation}
(n+1)\phi_\mathrm{min}=2\pi+\frac{2\pi(1-\epsilon)}{n+\epsilon}=\frac{2\pi(1-\epsilon)}{n+\epsilon}~\mathrm{(modulo~2\pi)}.
\end{equation}
Therefore $(n+1)\phi_\mathrm{min}<\phi_\mathrm{min}$ (modulo $2\pi$), and cannot belong to $\Phi$ unless $\epsilon=0$. One concludes that the set $\Phi_n$ of $n$ phase-shifts has to be of the form:
\begin{equation}
\Phi_n=\left\{ \phi_{n,p}=\left. \frac{2\pi p}{n} \right| ~~n\in  \mathrm{I\!N},~p\in \mathrm{I\!N},~0\leq p < n\right\},
\end{equation}
which finally is a quite obvious choice.
\end{appendix}

\end{document}